\documentclass[conference]{IEEEtran}

\IEEEoverridecommandlockouts
\usepackage{cite}
\usepackage{amsmath,amssymb,amsfonts}
\usepackage{algorithmic}
\usepackage{graphicx}
\usepackage{textcomp}
\usepackage{xcolor}
\usepackage[caption=false]{subfig}
\def\BibTeX{{\rm B\kern-.05em{\sc i\kern-.025em b}\kern-.08em
    T\kern-.1667em\lower.7ex\hbox{E}\kern-.125emX}}
\begin{document}

\title{High-resolution Coastline Extraction in SAR Images via MISP-GGD Superpixel Segmentation \\
\thanks{Funding for this work was provided in part by the Engineering and Physical Sciences Research Council (EPSRC) under grant EP/R009260/1 (AssenSAR) awarded to A. Achim, by the Royal Society via grants DHF\slash{R1}\slash{180068} and RGF\slash{EA}\slash{181086} awarded to B. Adams and by the EPSRC Impact Acceleration Account held by the University of Bristol via the Digital Environment: Using Machine Learning and Satellite data to Detect Movements along the Rail Corridor 2021-2342 grant awarded to N. Anantrasirichai.}
}

\author{\IEEEauthorblockN{Odysseas Pappas}
\IEEEauthorblockA{\textit{Visual Information Laboratory} \\
\textit{University of Bristol}\\
Bristol, UK \\
o.pappas@bristol.ac.uk}
\and
\IEEEauthorblockN{Nantheera Anantrasirichai}
\IEEEauthorblockA{\textit{Visual Information Laboratory} \\
\textit{University of Bristol}\\
Bristol, UK \\
n.anantrasirichai@bristol.ac.uk}
\and
\IEEEauthorblockN{Byron Adams}
\IEEEauthorblockA{\textit{School of Earth Sciences} \\
\textit{University of Bristol}\\
Bristol, UK \\
byron.adams@bristol.ac.uk}
\and
\IEEEauthorblockN{Alin Achim}
\IEEEauthorblockA{\textit{Visual Information Laboratory} \\
\textit{University of Bristol}\\
Bristol, UK\\
alin.achim@bristol.ac.uk}
}

\maketitle

\begin{abstract}
High accuracy coastline/shoreline extraction from SAR imagery is a crucial step in a number of maritime and coastal monitoring applications. We present a method based on image segmentation using the Generalised Gamma Mixture Model superpixel algorithm (MISP-GGD). MISP-GGD produces superpixels adhering with great accuracy to object edges in the image, such as the coastline. Unsupervised clustering of the generated superpixels according to textural and radiometric features allows for generation of a land/water mask from which a highly accurate coastline can be extracted. We present results of our proposed method on a number of SAR images of varying characteristics.
\end{abstract}


\section{Introduction}
Synthetic Aperture Radar (SAR) has become one of the most widely employed remote sensing modalities for a number of maritime and coastal monitoring tasks, largely due to its ability to produce highly detailed imagery regardless of the day/night cycle and meteorological conditions. Applications such as mapping, automated shipping navigation, ship-at-sea detection, coastal erosion monitoring, reservoir fault monitoring, glacial lake outburst flood monitoring and more often employ SAR data \cite{Buono2014} \cite{Pappas2018}. In many such applications the ability to automatically delineate water bodies and accurately extract the coastline/shoreline from a SAR image is of paramount importance.

The inherent speckle noise of SAR imagery, and the occasionally low contrast between sea and land in the case of high sea states (adverse weather) make this task non-trivial. Coastline detection algorithms typically involve a process for segmenting the image into two broad classes of water and land, and subsequent morphological processing for the extraction of the continuous single-pixel boundary between the two. A simple median filtering operation followed by Otsu's thresholding is used to derive a land/sea mask in \cite{Ji2016}, followed by morphological filling and dilation to extract a continuous coastline. In \cite{Liu2016} the authors use a modified K-means algorithm for image segmentation, followed by an object-based region merging algorithm to derive land and water masks. 

As the segmentation of the image into land/water regions is clearly the more challenging task involved, a number of methods have been proposed that take advantage of additional available information to inform this segmentation task. This includes the the multiple polarisations offered by many modern SAR platforms, or the coherence information that can be obtained from interferometric image pairs (InSAR).

In \cite{Dellepiane2004} for example the authors perform land/water segmentation on a texture-smoothed coherence image derived from an InSAR image pair. A CFAR detector whose threshold selection is informed by a multi-polarisation analysis of the sea surface is utilised in \cite{Buono2014}, to produce a binary image from which a coastline is extracted by means of a Sobel edge detector. In \cite{Nunziata2016} the authors use the a metric based on co- and cross-polarised channel correlation to perform land/water discrimination. 

We here propose an approach using the Generalised Gamma Mixture Model Superpixel algorithm (MISP-GGD) for image segmentation and unsupervised agglomerative clustering for the classification of the image segments into land and water regions based on their textural and radiometric properties, followed finally by morphological post-processing for coastline extraction.

The MISP-GGD algorithm produces superpixels with excellent adherence to object boundaries in SAR images \cite{Pappas2020}, allowing for accurate delineation of complex linear structures. As a consequence of the boundary adherence of MISP-GGD the generated superpixel tend to only include pixels of one class, hence making the classification/clustering step more straightforward. Additionally, this method requires no training data, can be performed on a single image with no requirement for interferometric image pairs or multipolarised products, is highly resilient to data variability (in terms of polarisation, incidence angle, sea state) and does not involve any filtering operations (as in e.g. \cite{Dellepiane2004} \cite{Ji2016}) that could introduce blurring of the coastline.

The remainder of this paper is organised as follows: Section 2 describes the MISP-GGD superpixel algorithm that underpins the proposed coastline extraction algorithm. Section 3 describes the remaining steps of the algorithm, including superpixel classification and land/sea mask post-processing. Results on SAR images from various platforms are presented in Section 4, while Section 5 provides some closing comments and conclusions.

\section{MISP-GGD Superpixel Segmentation}
Superpixel algorithms oversegment an image into homogeneous groupings of perceptually similar pixels, with the aim of utilising these segments as meaningful primitives for further image processing tasks such as object segmentation \& classification, target detection \& tracking and others, for a variety of image data types and applications \cite{Achanta2012}. 

As SAR images differ significantly from natural images, primarily in that they lack colour information and exhibit significant levels of speckle noise, most standard superpixel algorithms offer sub-optimal results. SAR-specific superpixels have therefore started appearing in the literature. These include the Mixture Model Superpixel algorithm (MISP)\cite{Arisoy2016} and its Generalised Gamma variant (MISP-GGD) \cite{Pappas2020} utilised here, which rely on a finite mixture model (FMM) modelling SAR image pixel amplitudes/intensities as well as their spatial coordinates. 

In MISP-GGD, the Generalised Gamma Distribution \cite{Li2011} is used to model SAR pixel amplitudes and/or intensities, while the spatial coordinates are modelled as a 2-D Gaussian distribution. Two pixels belonging to the same cluster (superpixel) are then assumed to have their amplitudes generated from the same Generalised Gamma distribution (i.e. the same mixture component in the FMM) and their spatial coordinates from the same Gaussian distribution. An iterative process is then employed to assign superpixel labels to each pixel in the image and to infer the related statistical parameters for the FMM.

We denote a pixel's amplitude $a_n$ and spatial coordinates $\textbf{q}_n = [x_n,y_n]^T$   where $n=[1,...,N]$ is the lexicographically ordered pixel index with $N$ being the total number of pixels in the image. The image is to be segmented into a total number of $K$ mutually exclusive superpixels, with the $k$th superpixel denoted as $S_k$ where $k\in[1,...,K]$. 

A pixel's amplitude $a_n$ is therefore modelled as

\begin{equation}
p(a_n) = \frac{|v_k|\kappa_k^{\kappa_k}}{\sigma_k\Gamma(\kappa_k)}\left(\frac{a_n}{\sigma_k}\right)^{\kappa_k v_k -1}\exp\left\lbrace-\kappa_k\left(\frac{a_n}{\sigma_k}\right)^{v_k}\right\rbrace, 
\label{eq:ggd_pdf2}
\end{equation}

\noindent where $v_k, \kappa_k, \sigma_k$ are the GGD power, shape and scale parameters respectively that model the $k$th superpixel.

The spatial distribution of pixels around a superpixel centroid is assumed to be normal and is given as follows

\begin{equation}
p(\textbf{q}_n) = \frac{1}{\sqrt{2\pi\Sigma_k}}\exp\left\lbrace -\frac{1}{2}(\textbf{q}_n-\textbf{m}_k)^T \Sigma_k^{-1}(\textbf{q}_n-\textbf{m}_k)\right\rbrace,
\label{eq:2dgauss_pdf}
\end{equation}

\noindent where $\textbf{m}_k$ is the $k$th superpixel's centroid spatial coordinates and $\Sigma_k$ is the covariance matrix of that same superpixel, leading to a complete feature set of $\theta_k = \left\lbrace v_k, \kappa_k, \sigma_k, \textbf{m}_k,\Sigma_k\right\rbrace$,. 

Each pixel in the image is assigned a label in the form of a $K$-dimensional binary vector $\textbf{z}_n\in\left\lbrace[1,0,...,0],[0,1,...,0],...,[0,0,...,1]\right\rbrace$ so that $\sum_{k=1}^K \textbf{z}_{n,k} = 1$. The natural prior for $\textbf{z}_n$ is a multinomial distribution and is defined as $p(z_n | \omega_{1:K}) = \prod_{k=1}^K \omega_k^{z_{n,k}}$ where $\omega_{1:K}$ are the parameters of the multinomial distribution. 

Under conditional probability rules, the joint density of a pixel's feature vector $\textbf{f}_n$ and label vector $\textbf{z}_n$ can be equated to

\begin{equation}
p(\textbf{f}_n,\textbf{z}_n|\theta_{1:K},\omega_{1:K}) = p(\textbf{f}_n|\textbf{z}_n,\theta_{1:K})p(\textbf{z}_n|\omega_{1:K}),
\label{eq:joint1}
\end{equation}

\noindent and it can be further shown \cite{Arisoy2016} that the following finite mixture density can be obtained

\begin{equation}
p(\textbf{f}_n|\theta_{1:K},\omega_{1:K}) = \sum_{\textbf{z}_n} \prod_{k=1}^K\left[p(f_n|\theta_k)\omega_k\right]^{z_{n,k}},
\label{eq:joint2}
\end{equation}

\noindent where $\omega_k$ corresponds to the mixture proportion of the superpixels. A conjugate Dirichlet prior for these proportions can then be defined as 

\begin{equation}
p(\omega_{1:K}) = \frac{1}{B(\alpha)}\prod_{k=1}^K\omega_k^{\alpha-1},
\label{eq:diriprior}
\end{equation}

\noindent where $\alpha$ is the concentration parameter and $B(\alpha) = \Gamma^K(\alpha) / \Gamma(\alpha K)$, where $\Gamma(\cdot)$ is the Gamma function.

The parameter set $\theta_{1:K}$, the mixture proportion $\omega_{1:K}$ and the superpixel label $\textbf{z}_{1:N}$ of each pixel are estimated using the block iterated conditional mode algorithm (ICM) \cite{Besag1974}. This allows the variables to be updated along iterations as:

\begin{equation}
\textbf{z}_n^t \leftarrow \max \: p(\textbf{f}_n|\textbf{z}_n,\theta_{1:K}^{t-1})p(\textbf{z}_n|\omega_{1:K}^{t-1}),
\label{eq:upd1}
\end{equation}
\begin{equation}
\theta_k^t \leftarrow \max \: p(\textbf{f}_{1:N}|\textbf{z}_{1:N},\theta_k),
\label{eq:upd2}
\end{equation}
\begin{equation}
\omega_k^t \leftarrow \max \: p(\textbf{z}_{1:N}^t|\omega_{1:K})p(\omega_{1:K}),
\label{eq:upd3}
\end{equation}

\noindent where $t$ is the iteration index. From these the mixture proportion can be worked out as

\begin{equation}
\omega_k^t = \frac{\sum_{n=1}^Nz_{n,k}^{t-1}+\alpha-1}{N + K(\alpha-1)},
\label{eq:upd_omega}
\end{equation}

For more details on MISP-GGD, including on closed-form parameter estimation for the GGD performed according to the second-kind cumulants method, see \cite{Pappas2020}.

MISP-GGD offers superior boundary adherence for superpixel segmentation of SAR images, and is capable of dealing with a very wide gamut of image specifications, including both amplitude and intensity image products of various polarisations and incidence angles.

\section{Coastline Extraction}
For the purposes of this application all pixels in an image can be considered as belonging to one of only two categories, landmass and water (open sea or otherwise). Courtesy of the properties of the MISP-GGD algorithm, superpixels along the land/water boundary (i.e. the coastline/shoreline) have their edges align well with the image object edges. It is furthermore expected that each superpixel will contain only one of the two classes of pixels, as a direct consequence of this boundary adherence property \cite{Pappas2020}. Segmenting therefore the image into landmass and water regions on the basis of a MISP-GGD superpixel over-segmentation can yield an accurate delineation of the coastline.

The process therefore consists of superpixel segmentation, classification of superpixels into land/water cover, outlier elimination and finally coastline extraction. Superpixel segmentation is performed with the MISP-GGD algorithm outlined above. The classification task must then be addressed; this task is relatively straightforward, given we are here only interested in two broad classes of which the members differ dramatically. 

Landmass superpixels and water superpixels differ significantly in both their radiometric and textural properties, with water cover typically exhibiting a lot less texture and darker radar returns compared to land. We use the information entropy as well as the median value of the pixels enclosed in each superpixel as features for characterising that superpixel, as they can be directly indicative of textural and radiometric dissimilarity respectfully. 

Recall that for a random population $X$ with possible values $\lbrace x_1,x_2,...,x_i,...x_n \rbrace$ the entropy value $S(X)$ is given by

\begin{equation}
S(X) = - \sum_i p(x_i) \log p(x_i)
\label{eq:entropy}
\end{equation}

On the basis of these two features, superpixels are classified into two clusters in a non-supervised agglomerative fashion. Of these two clusters one will therefore correspond to the landmass and one to the water-containing superpixels. While it is possible to include more features, as in e.g. \cite{Pappas2020}, we have found this to not be necessary in this case. As mentioned above, the good adherence of the MISP-GGD superpixels to the coast boundary actually helps simplify the classification task by keeping the number of actual inter-class superpixels to a minimum (ideally zero).

\begin{figure}[t]
\centering
\subfloat[]{\includegraphics[width=0.23\textwidth]{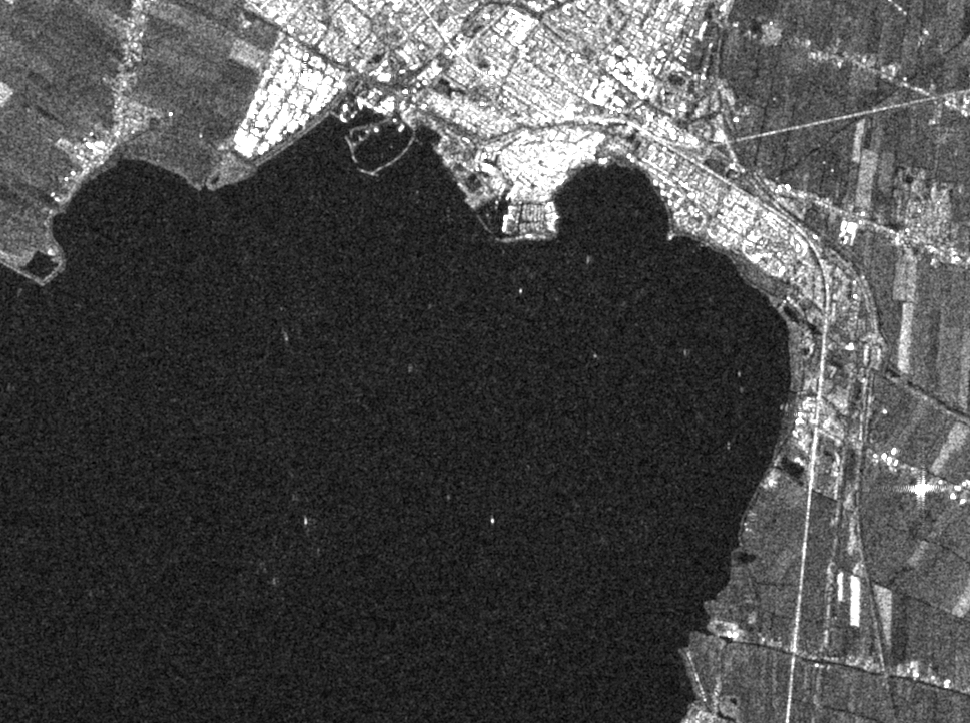}}
\hspace{0.05cm}
\subfloat[]{\includegraphics[width=0.23\textwidth]{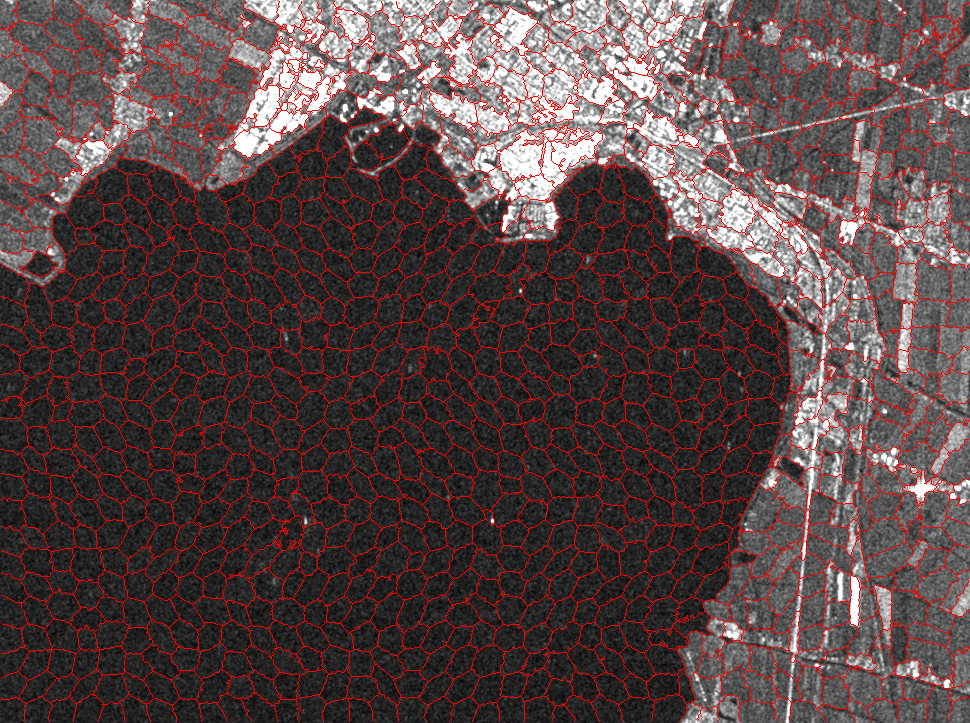}}

\subfloat[]{\includegraphics[width=0.23\textwidth]{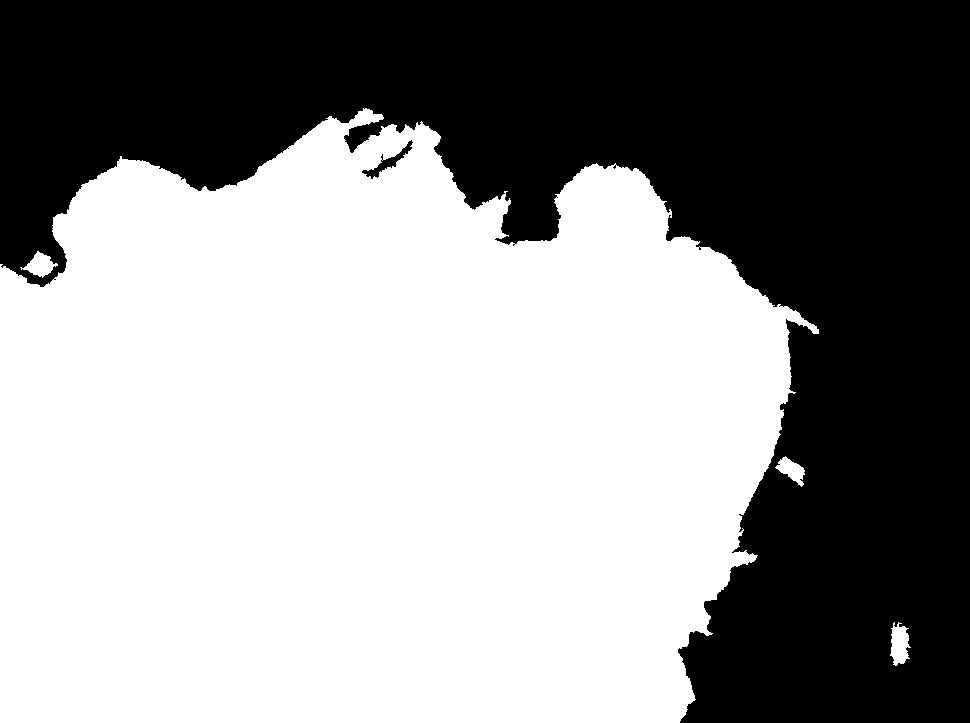}}
\hspace{0.05cm}
\subfloat[]{\includegraphics[width=0.23\textwidth]{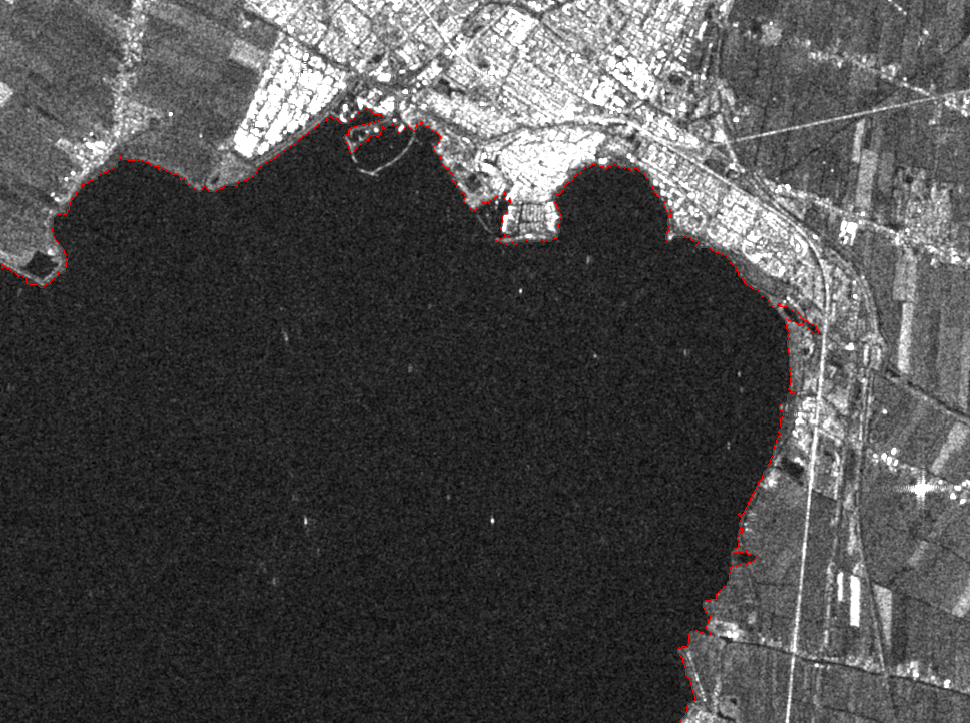}}
\caption{Coastline detection on SENTINEL-1 IW image. (a) Original image crop, (b) MISP-GGD superpixel segmentation, (c) Land/sea classification mask (prior to filling), (d) Extracted coastline.}
\label{Sentinel 1}
\end{figure}

There can of course be cases of a superpixel being misclassified. The performance of MISP-GGD superpixels is such that they very rarely contain pixels from both classes; they tend to contain purely land or sea as the superpixels follow the image object edges and tend to only contain pixels from one object/area. As a result we rarely have a superpixel missclassified due to containing pixels from both classes. However, one might have superpixels within the land region classified as water (e.g. due to in-land water bodies such as small lakes) or superpixels within the sea region classifies as land members (e.g. due to small islands). These can be very easily eliminated via a void-filling operation on the connected component map that is produced by the previous agglomerative clustering step, so as to produce a final binary map of only two components, the interface of which corresponds to the coastline. 

Finally, extraction of this single pixel wide coastline can be achieved via a border operation on any of the two connected components, excluding of course the parts of the border that correspond to the actual image borders.

\section{Experimental Results}

\begin{figure}[t]
\centering
\subfloat[]{\includegraphics[width=0.23\textwidth]{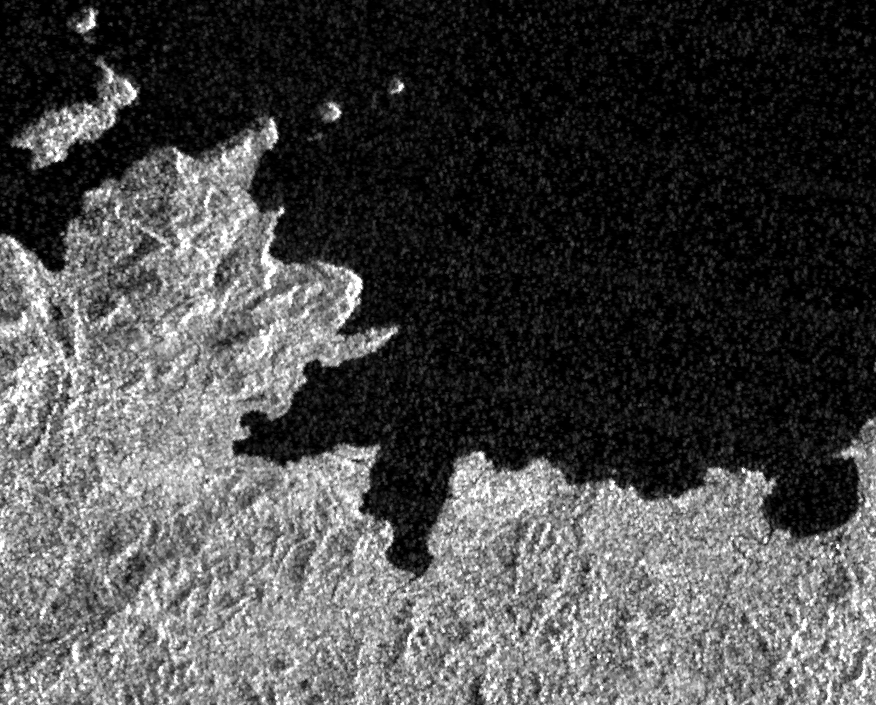}}
\hspace{0.05cm}
\subfloat[]{\includegraphics[width=0.23\textwidth]{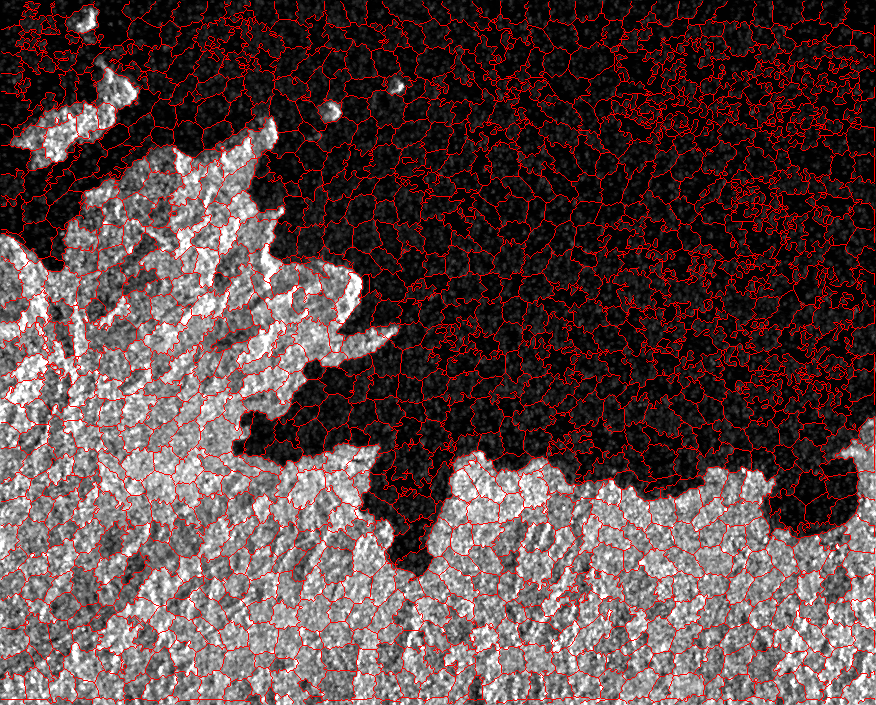}}

\subfloat[]{\includegraphics[width=0.23\textwidth]{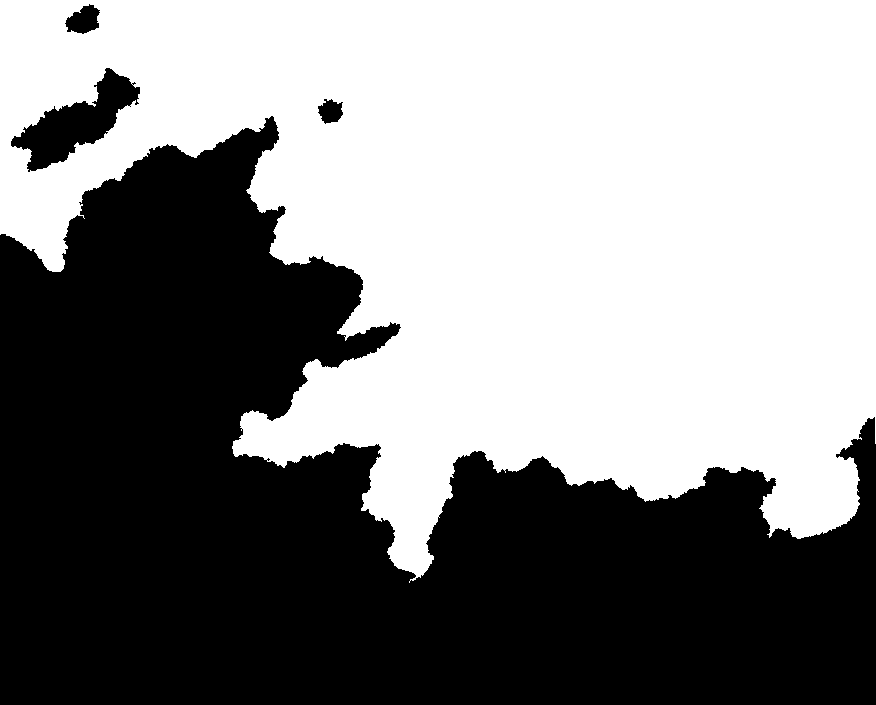}}
\hspace{0.05cm}
\subfloat[]{\includegraphics[width=0.23\textwidth]{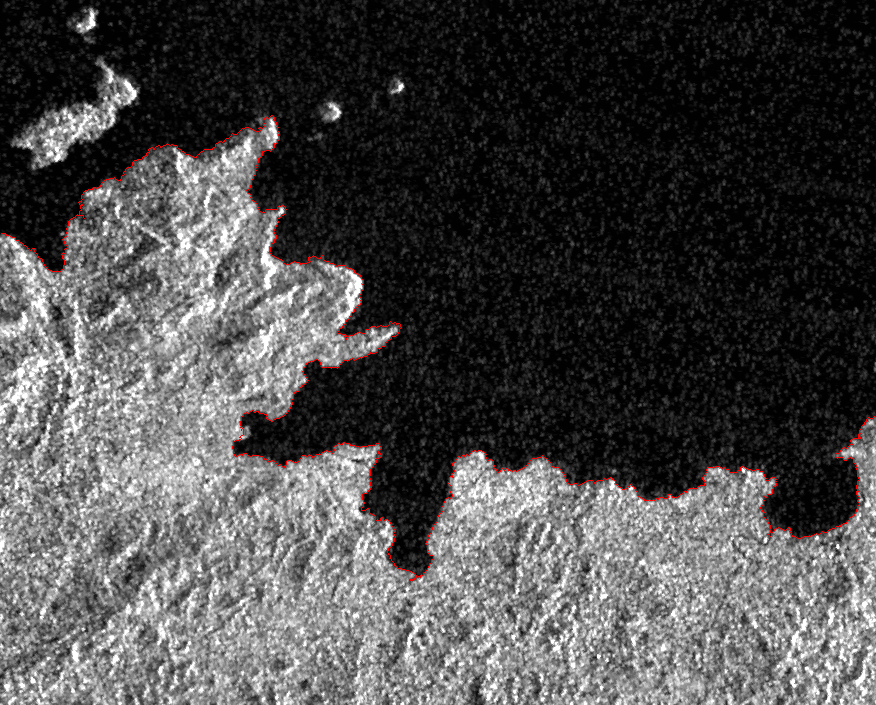}}
\caption{Coastline detection on TerraSAR-X WS image. (a) Original image crop, (b) MISP-GGD superpixel segmentation, (c) Land/sea classification mask (prior to filling), (d) Extracted coastline.}
\label{TerraSAR-X}
\end{figure}

We present results of our proposed method on a number of SAR images. Image 1 (Fig. \ref{Sentinel 1}) is a crop (723x970 pixels) of a SENTINEL-1 Interferometric Wide Swath (IW), Ground Range Detected (GRD) Amplitude image in VH polarisation, showing the area near Hoorn in the Netherlands at a spatial resolution of 10m. 

Image 2 (Fig. \ref{TerraSAR-X}) is a crop (705x876 pixels) of a TerraSAR-X Wide Scan (WS), Multi Look Ground Range Detected (MGD) radar brightness $\beta$ image in HV polarisation, showing part of the coast of Panama at a spatial resolution of 40m.

\begin{figure}[t]
\centering
\subfloat[]{\includegraphics[width=0.23\textwidth]{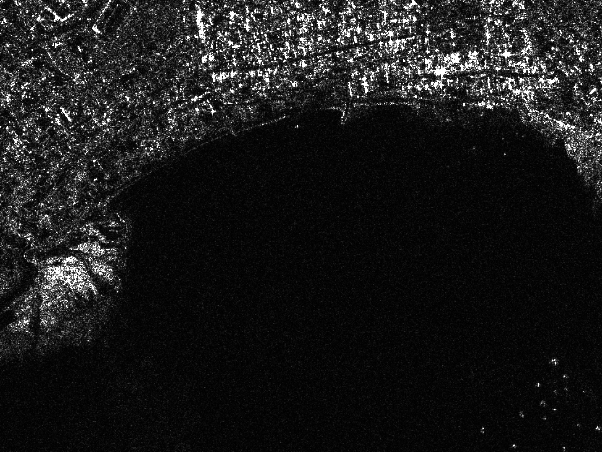}}
\hspace{0.05cm}
\subfloat[]{\includegraphics[width=0.23\textwidth]{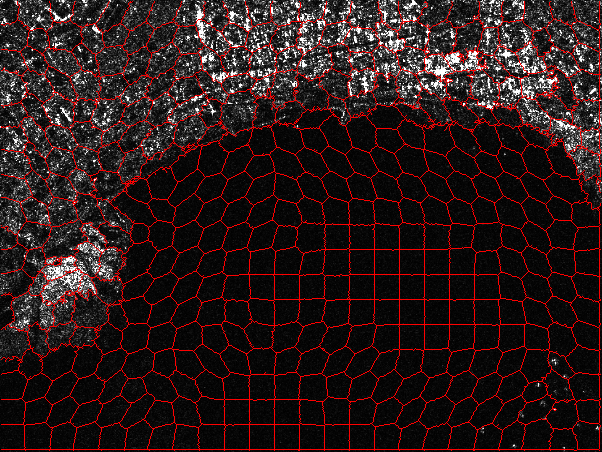}}

\subfloat[]{\includegraphics[width=0.23\textwidth]{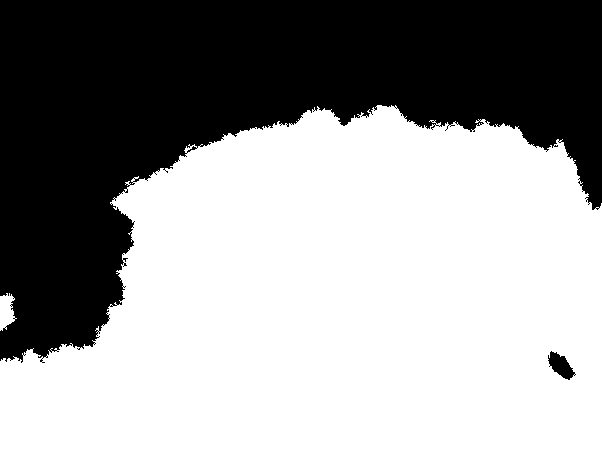}}
\hspace{0.05cm}
\subfloat[]{\includegraphics[width=0.23\textwidth]{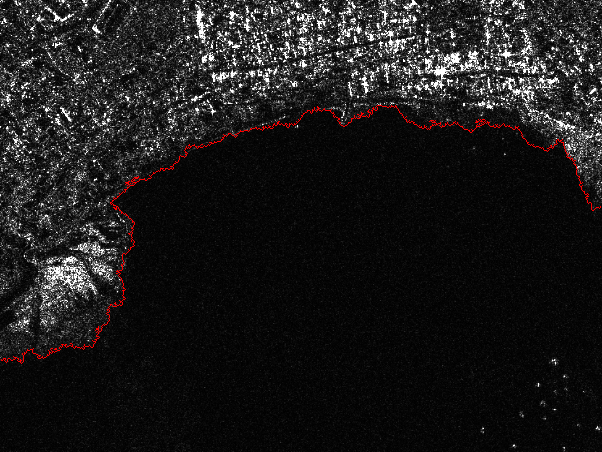}}
\caption{Coastline detection on ICEYE-X2 Stripmap image. (a) Original image crop, (b) MISP-GGD superpixel segmentation, (c) Land/sea classification mask (prior to filling), (d) Extracted coastline.}
\label{ICEYE-X2}
\end{figure}

Image 3 (Fig. \ref{ICEYE-X2}) is a crop (452x502 pixels) of an ICEYE-X2 Stripmap, GRD Intensity image in VV polarisation, showing part of the coast of Dar es Salaam, Tanzania, at a spatial resolution of 3m.

Figure \ref{Sentinel 1} shows an example of the proposed algorithm being utilised to detect the coastline on a SENTINEL-1 image. This particular section of coastline includes a few man-made structures and a couple of small harbors. The superpixel segmentation map clearly demonstrates the ability of MISP-GGD superpixels to accurately align their edges with the dominant edges of objects in the image, in this case the coastline. The binary map also shows a couple of misclassifications of the types discussed previously (lagoons, small islets, piers etc) which are easily eliminated by the morphological filling operation. The coastline is delineated in red in the final image. 

Figure \ref{TerraSAR-X} contains a more natural coastline with a variety of coves and small islands. The proposed method is again capable of producing a coastline which according to visual inspection seems to follow the natural coastline with great accuracy. Note that as seen in Fig \ref{TerraSAR-X} (c) the polarity of the binary land/water mask is irrelevant, as the process will produce a coastline regardless of whether the treated connected component corresponds to the land mask or the water mask.

Figure \ref{ICEYE-X2} shows again a similar set of examples only this time the presented image crop is taken from an ICEYE-X2 product. This being one of the most recent SAR satellites in orbit, it produces stripmap products at an astonishing resolution of only 3 meters, allowing for the delineation of very fine coastal features. The proposed method is again capable of accurately delineating this coastline, even given the comparatively weak contrast relationship between land and sea in this image.

\section{Conclusions}
In this paper we have presented a method for coastline/shoreline extraction in SAR images using the MISP-GGD SAR superpixel algorithm, followed by agglomerative clustering and morphological processing. The satisfactory performance of MISP-GGD in terms of boundary adhesion allows for the extraction of coastlines with good accuracy, as indicated by the presented results over a variety of SAR images. 

MATLAB code for implementing the MISP-GGD has been made publicly available as part of our previous work and can be found in the following Github repository: 

\noindent $https://github.com/odispap/SARSuperpixelRivers$. 

%


\bibliographystyle{IEEEtran}
\bibliography{coasts_biblio}

\end{document}